\documentclass[aps,prl,groupedaddress,superscriptaddress,twocolumn, longbibliography,floatfix]{revtex4-2}

\usepackage[utf8x]{inputenc}

\usepackage{amsfonts,amsmath,amssymb,stmaryrd}

\usepackage{graphicx}
\usepackage{hyperref}
\usepackage{mathrsfs}
\usepackage{siunitx}
\usepackage{color}
\usepackage{csquotes}
\usepackage[all]{hypcap}
\usepackage{braket}
\usepackage{wrapfig}
\usepackage{array}
\usepackage{upgreek}
\usepackage{subfigure}
\usepackage{comment}

\DeclareSIUnit{\erec}{E_\mathrm{rec}}

\begin{document}

\title{Experimental observation of a dissipative phase transition in a multi-mode many-body quantum system}

\author{J. Benary}
\affiliation{Department of Physics and Research Center OPTIMAS, Erwin-Schrödinger-Straße 46, Technische Universität Kaiserslautern, 67663 Kaiserslautern, Germany}

\author{C. Baals}
\affiliation{Department of Physics and Research Center OPTIMAS, Erwin-Schrödinger-Straße 46, Technische Universität Kaiserslautern, 67663 Kaiserslautern, Germany}

\author{E. Bernhart}
\affiliation{Department of Physics and Research Center OPTIMAS, Erwin-Schrödinger-Straße 46, Technische Universität Kaiserslautern, 67663 Kaiserslautern, Germany}

\author{J. Jiang}
\affiliation{Department of Physics and Research Center OPTIMAS, Erwin-Schrödinger-Straße 46, Technische Universität Kaiserslautern, 67663 Kaiserslautern, Germany}

\author{M. R\"ohrle}
\affiliation{Department of Physics and Research Center OPTIMAS, Erwin-Schrödinger-Straße 46, Technische Universität Kaiserslautern, 67663 Kaiserslautern, Germany}

\author{H. Ott}
\email{ott@physik.uni-kl.de}
\affiliation{Department of Physics and Research Center OPTIMAS, Erwin-Schrödinger-Straße 46, Technische Universität Kaiserslautern, 67663 Kaiserslautern, Germany}

\begin{abstract}

\noindent\textbf{Abstract:}
We characterize the dissipative phase transition in a driven dissipative Bose-Einstein condensate of neutral atoms. Our results generalize the work on dissipative nonlinear Kerr resonators towards many modes and stronger interactions. We measure the effective Liouvillian gap and analyze the microscopic system dynamics, where we identify a non-equilibrium condensation process.

\end{abstract}

\date{\today}

\maketitle

Dissipative phase transitions \cite{Minganti2018,Kessler2012,Boite2013,Vicentini2018,Walker2018,Fink2018} are characteristic of open systems, where the coupling to an external reservoir provides a source of dissipation which competes with an internal or external pumping process. Depending on the relative strength of the two processes, different steady states emerge or even coexist. Most notably, the steady-states can show properties which cannot be observed in equilibrium \cite{Wachtel2016}. The generic properties of first and second order dissipative phase transitions can be derived from a spectral analysis of the system's Liouvillian \cite{Kessler2012,Minganti2018}. Critical behavior at the phase transition has been studied both, theoretically \cite{Kessler2012,Marcuzzi2014,Casteels2016,Overbeck2017,Casteels2017,Vicentini2018,Letscher2017} and experimentally \cite{Letscher2017,Trenkwalder2016,Carr2013,Fink2017,Fink2018,Fitzpatrick2017,Rodriguez2017,Geng2020,Sahoo2020,li2021}.

One of the paradigmatic examples exhibiting a dissipative phase transition is a single mode quantum system, realized, e.g., with a nonlinear optical medium in a cavity, which is coherently pumped by a laser and subject to cavity losses \cite{Drummond1980,Fink2017,Fitzpatrick2017,Rodriguez2017,Fink2018}. The Hamiltonian part of such a Kerr type system reads

\begin{equation}
    \hat{H}=\Delta \hat{a}^\dagger\hat{a} + U\hat{a}^\dagger  \hat{a}^\dagger \hat{a}  \hat{a} +  F^*\hat{a} + F \hat{a}^\dagger,
    \label{master}
\end{equation}

where $\Delta$ is the detuning between the mode energy and the photon energy of the pump laser, $\hat{a}$ is the annihilation operator of the mode, $F$ is the amplitude of the pumping laser and $U$ denotes the interaction. The cavity losses are included via a master equation in Lindblad form,

\begin{equation}
\begin{aligned}
\dot{\hat{\rho}}&=\hat{\mathcal{L}}\hat{\rho}\\
&=-\frac{i}{\hbar}\left[\hat{H},\hat{\rho}\right]+\frac{\gamma_{\mathrm{diss}}}{2} \left(2\hat{a} \hat{\rho} \hat{a}^\dagger -  \hat{a}^\dagger\hat{a} \hat{\rho} - \hat{\rho}\hat{a}^\dagger\hat{a} \right),
\end{aligned}
\label{master2}
\end{equation}

where $\gamma_{\mathrm{diss}}$ is the loss rate and $\hat{\rho}$ is the density operator. This model exhibits several prominent features:\newline (i) There are two classes of states (with low and high occupation number) which are connected by a first order dissipative phase transition. (ii) While a mean-field treatment predicts optical bistability, the full quantum description (Eqs.$\,$\eqref{master},\eqref{master2}) allows for tunneling between the steady states, thus erasing bistability and rendering the states metastable. (iii) As a consequence of (i) and (ii), single trajectories show a pronounced switching behavior between the two classes of states \cite{Fink2018}. 

\begin{figure}[t]
\centering
\includegraphics[]{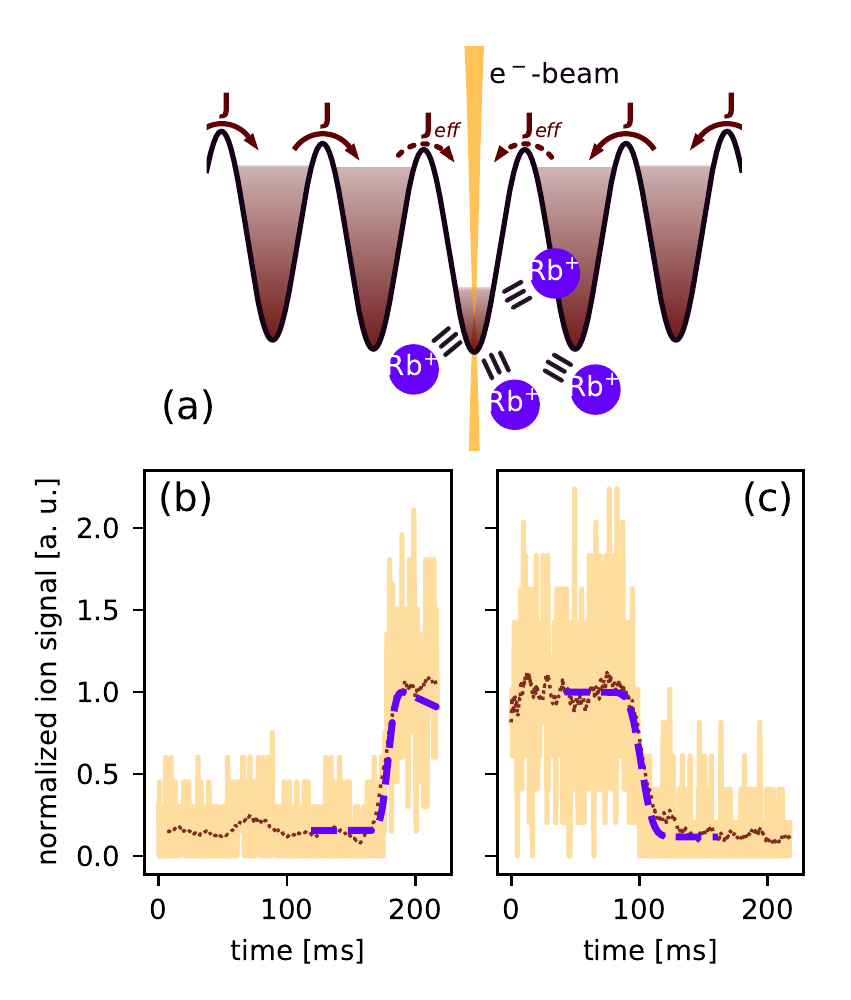}
\caption{(a) A Bose-Einstein condensate is loaded into a one-dimensional optical lattice. The individual sites of the condensate array are coupled via the coherent tunneling coupling $J$. The central site is subjected to particle loss, realized with an electron beam. The nonlinear interaction leads to an effective tunneling coupling $J_{\mathrm{eff}}$ to the neighboring sites (see text). The produced ions are guided towards a detector and provide a time-resolved measurement signal. (b) Single experimental trajectory (amber line, moving average is shown as dotted line) for an initial state with low atom number in the central site. After $\SI{200}{m\s}$ evolution time, the trajectory switches to a high atom number, which is comparable to that of the neighboring sites. This jump is fitted with a sigmoid function (indigo dashed line). (c) Same as (b) for an initially high atom number in the central site.}
\label{setup}
\end{figure}

For multi-mode systems instead, much less is known. There exist only a few theoretical works \cite{Dagvadorj2015,Boite2013,Vicentini2018,Foss-Feig2017,Reeves2021,Mink2022}, mostly on the driven-dissipative 2D Bose-Hubbard model. The extension to many modes in the interacting many-body regime poses many challenges to its theoretical description (see methods for an explicit form of the master equation). In practice, already a few tens of modes and particles make exact diagonalization intractable. Experimental studies of a dissipative phase transition in a multi-mode system therefore not only generalizes the results of the single mode Kerr oscillator, but also provide benchmarks for advanced theoretical methods to describe driven-dissipative systems. 

In this work, we investigate the multi-mode version of the nonlinear Kerr resonator in an atomic quantum gas experiment. First results for the system under investigation have been reported previously by one of the authors \cite{NDC,Bistability}. We present experimental indications that this system undergoes a first order dissipative phase transition when tuning the ratio between the coherent drive and the dissipation. We observe the same characteristic switching behavior as in the single mode system and find signatures of a closing of the Liouvillian gap. On intermediate timescales, the system exhibits hysteresis and metastability. In particular, we identify and analyze a non-equilibrium condensation process. The system has triggered recent theoretical work using c-field methods \cite{Reeves2021} and a single-mode variational truncated Wigner approach \cite{Mink2022}.

The experimental system (Fig.$\,$\ref{setup}a) consists of a Bose-Einstein condensate loaded into a 1D optical lattice. This can be considered as an array of coupled two-dimensional Bose-Einstein condensates of neutral atoms. Each condensate occupies about $\SI{80}{}$ transverse harmonic oscillator modes. With a total of about $\SI{800}{}$ atoms in each condensate, the system is deep in the multi-mode regime with high mode occupation numbers. The central lattice site is subject to a homogeneous loss process with rate $\gamma_{\mathrm{diss}}$. The loss mechanism is realized by a focused electron beam, which is scanned across the central site and removes the atoms. A fraction of the atoms is converted into ions \cite{SEM}, which form the basis of our measurement signal. This way, we can measure the time-resolved atom number and the transverse extent of the atomic cloud in the central site. The lattice sites are connected via quantum mechanical tunneling with rate $J$, which constitutes the drive into the central site. The experimental sequence starts by initially preparing the system either with an ''empty'' site (low occupation number in the central site) or with a ''full'' site (same occupation number in all sites). We then record the temporal evolution of the site occupancy under the simultaneous influence of drive and dissipation.

In Fig.$\,$\ref{setup}, we show typical individual trajectories observed in the experiment for an initially empty (Fig.$\,$\ref{setup}b) and full central site (Fig.$\,$\ref{setup}c). The trajectories are characterized by a switching behavior, indicated by pronounced jumps in the site occupancy between the two states. The absolute magnitude of these jumps is large: about $\SI{700}{}$ atoms tunnel within a short time into or are removed from the central site. This switching behavior is reminiscent of that of the corresponding single mode system \cite{Fink2018} and hints at a dominant role of fluctuations. It has also been seen in a c-field treatment of this system \cite{Reeves2021}.

The nature of the appearing states between which the system switches has been identified previously \cite{Bistability,CPA}. For high occupation number, when the coherent drive dominates over the dissipation, there is a condensate in the central site, which is phase locked to the neighboring condensates. The whole system is in a superfluid state, all sites have similar atom numbers and the system exhibits steady-state transport known as coherent perfect absorption \cite{CPA}. In the opposite limit of large loss rate, the central site features only a small occupation in each mode. The atomic ensemble is in a normal state and macroscopic phase coherence is absent. Hence, superfluid transport into that site is inhibited \cite{Bistability}. In the vicinity of the phase transition, where dissipation and drive are of similar strength,  jumps are induced between the states.

To quantify the switching dynamics, we measure the time stamps of the jump for each trajectory. We then determine the time scales $\tau_{\mathrm{normal \rightarrow superfluid}} = \tau_{\mathrm{ns}}$ and $\tau_{\mathrm{superfluid \rightarrow normal}} = \tau_{\mathrm{sn}}$ for switching between the two configurations as well as the corresponding rates $\gamma_{\mathrm{ns}} = \tau_{\mathrm{ns}}^{-1}$ and $\gamma_{\mathrm{sn}} = \tau_{\mathrm{sn}}^{-1}$. The physical meaning of the switching rates becomes clear if one looks at the spectral decomposition of the Liouvillian
\begin{equation}
\hat{\mathcal{L}}\hat{\rho_i}=\lambda_i \hat{\rho}_i.
\label{eq:spectrum}
\end{equation}
Here, $\lambda_i$ denotes the complex eigenvalue, and $\operatorname{Re}(\lambda_i)\leq 0$ is the damping rate of $\hat{\rho}_i$. Eigenvalues with $\operatorname{Re}(\lambda_i)= 0$  denote steady states. Note that for any non-zero $\operatorname{Re}(\lambda_i)$, the $\hat{\rho}_i$ are not proper density matrices, as the trace is not preserved in time. A proper density matrix therefore always contains a contribution from at least one steady state. If two eigenvalues of Eq.$\,$\eqref{master} are zero, the system exhibits true bistability. If only one eigenvalue is zero, the spectrum of Eq.$\,$\eqref{eq:spectrum} shows a gap, the Liouvillian gap, which is defined by the smallest nonzero eigenvalue, $\min (-\operatorname{Re}(\lambda_i))$. The Liouvillian gap therefore also defines the slowest relaxation rate of the system.

As the full many-body problem is intractable, we adapt a simplified toy model, which was developed in \cite{Wilson2016} for the driven dissipative Bose-Hubbard model and applied in \cite{Fitzpatrick2017}. The toy model considers only two states between which the system can switch: the normal state and the superfluid state, represented by the density matrices $\rho_\mathrm{s}$ and $\rho_\mathrm{n}$. The corresponding rate equation model features two steady states. One is a stable steady-state with eigenvalue $\lambda=0$: $\rho_{\mathrm{ss}}=(\gamma_{\mathrm{ns}}\rho_{\mathrm{s}}+\gamma_{\mathrm{sn}}\rho_{\mathrm{n}})$. The second one is an exponentially decaying state, $\rho_{\mathrm{dec}}=(\gamma_{\mathrm{ns}}\rho_{\mathrm{s}}-\gamma_{\mathrm{sn}}\rho_{\mathrm{n}})$, with eigenvalue $\lambda_\mathrm{gap}=-(\gamma_{\mathrm{ns}}+\gamma_\mathrm{sn})$. Hence, the toy model features an effective Liouvillian gap, given by $\lambda_\mathrm{gap}$. Because the steady-state $\rho_{\mathrm{ss}}$ is a mixture of the normal and the superfluid state, a time resolved measurement of the site occupancy is characterized by jumps between the two states.

\begin{figure}[b]
\centering
\includegraphics[]{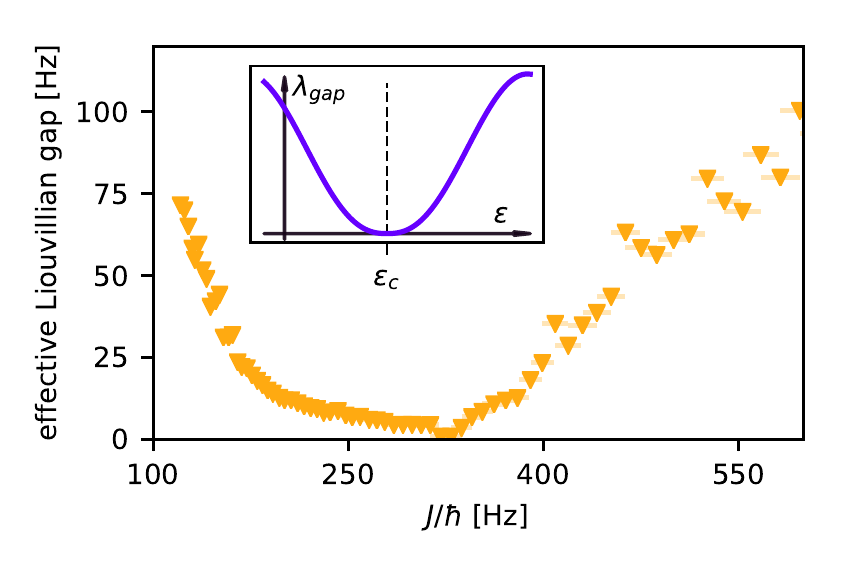}
\caption{Effective Liouvillian gap. We plot $\gamma_{\mathrm{ns}}+\gamma_{\mathrm{sn}}$ for a dissipation rate of $\gamma_{\mathrm{diss}} = \SI{250}{Hz}$ as a function of tunneling coupling $J/\hbar$. Inset: Paradigm of first-order dissipative phase transition. As the parameter $\varepsilon$ reaches the critical value $\varepsilon_c$ the Liouvillian gap $\lambda_{\mathrm{gap}}$ closes in the thermodynamic limit. Metastability is found in the vicinity of the critical point, giving rise to hysteresis.}
\label{rates}
\end{figure}

\begin{figure}[t]
\centering
\includegraphics[]{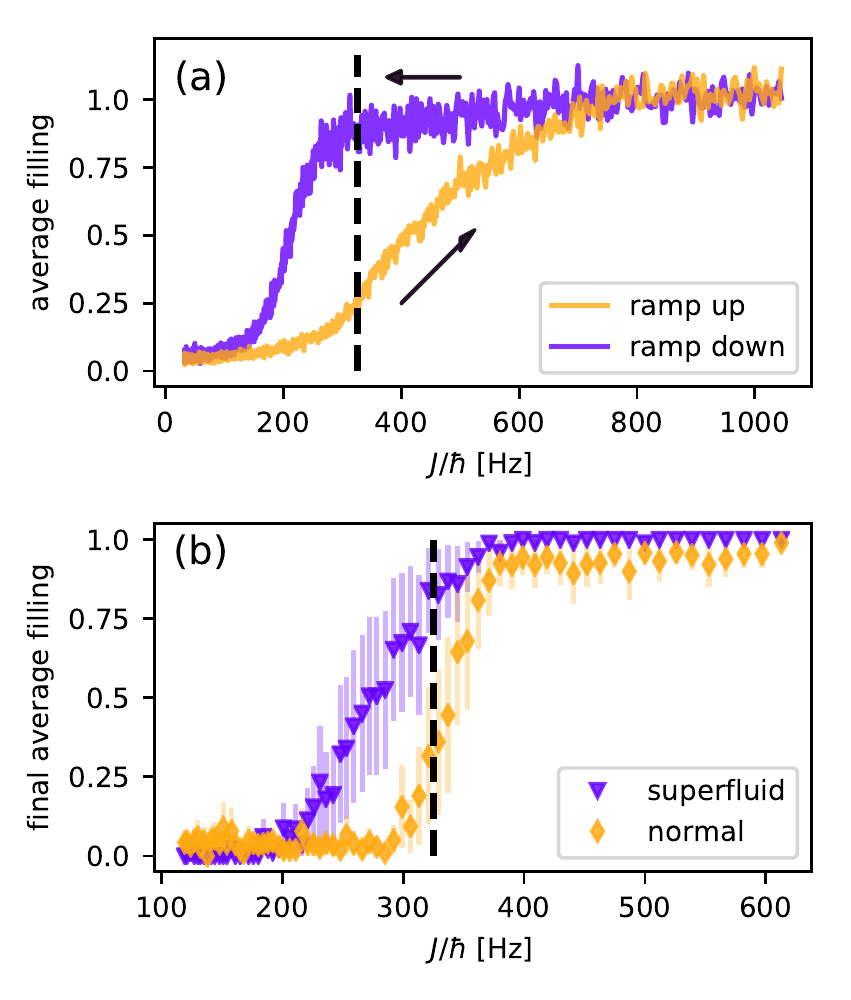}
\caption{(a) Dynamic hysteresis. The amber and indigo line represent the measured ion signal when the tunneling coupling is ramped up and back down, respectively, while keeping the dissipation strength constant at $\gamma_{\mathrm{diss}}=\SI{250}{\Hz}$. The dynamic hysteresis curve shown is averaged over 130 realizations. Within the total ramp duration $t_{\mathrm{ramp}}=\SI{650}{\ms}$ the tunneling coupling is changed between $J/\hbar = \SI{30}{\Hz}$ and $J/\hbar = \SI{1040}{\Hz}$ and back. (b) Metastability. We show the average filling after a evolution time of $t_{\mathrm{meas}}=\SI{200}{\ms}$ for the two different initial conditions ($\gamma_{\mathrm{diss}}=\SI{250}{\Hz}$). The black dashed lines indicate the phase transition extracted from Fig.$\,$\ref{rates}.}
\label{jumpshare}
\end{figure}

\begin{figure}[t]
\centering
\includegraphics[]{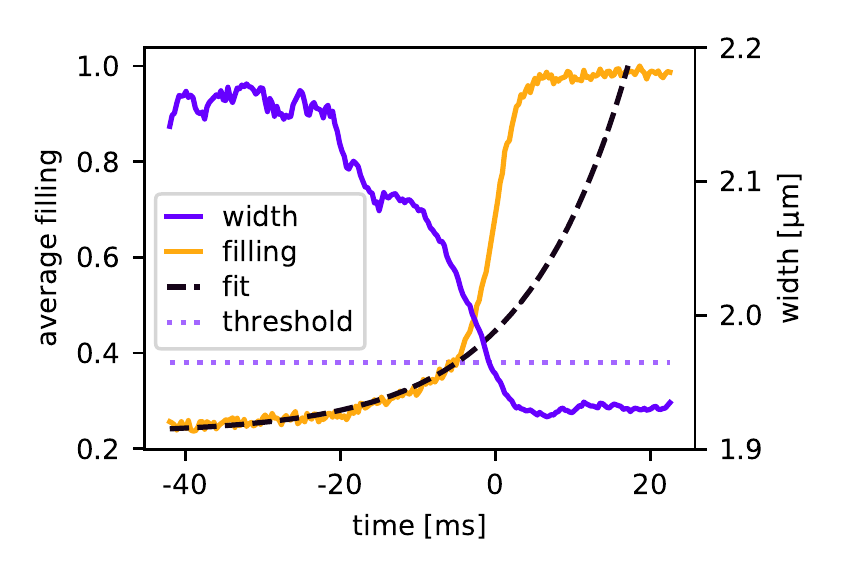}
\caption{Average jump. The average atom number (amber) and width of atomic cloud (indigo) during a state change are plotted. An initial exponential growth of the atom number (black dashed line) is abruptly sped up above a certain threshold filling. At the same time, the transverse extent shrinks.}
\label{avgJump}
\end{figure}

\begin{figure}[b]
\centering
\includegraphics[]{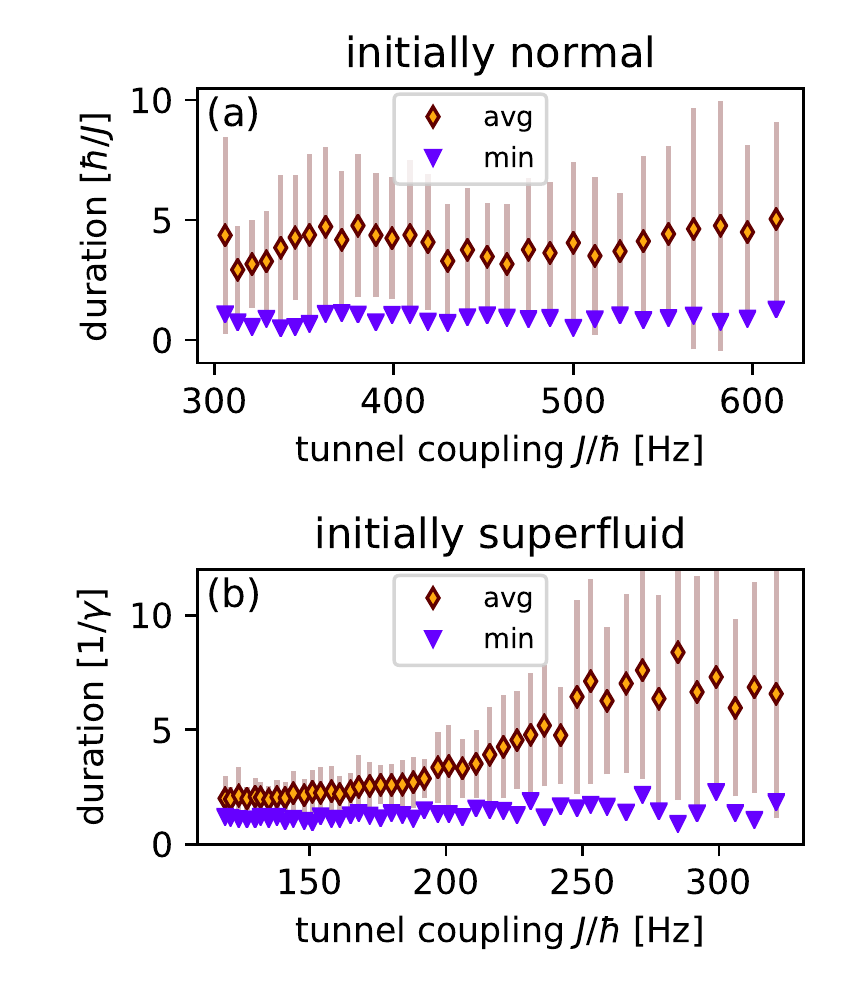}
\caption{Duration of jumps. (a) The minimum duration of the jumps from the normal to the superfluid state is determined by the tunneling coupling $J/\hbar$. On average the jumps take about four tunneling times. (b) For state changes from the superfluid to the normal state we find that the minimum duration is determined by the inverse of the dissipation rate $\gamma_{\mathrm{diss}}$.}
\label{duration}
\end{figure}

\begin{figure}[t]
\centering
\includegraphics[]{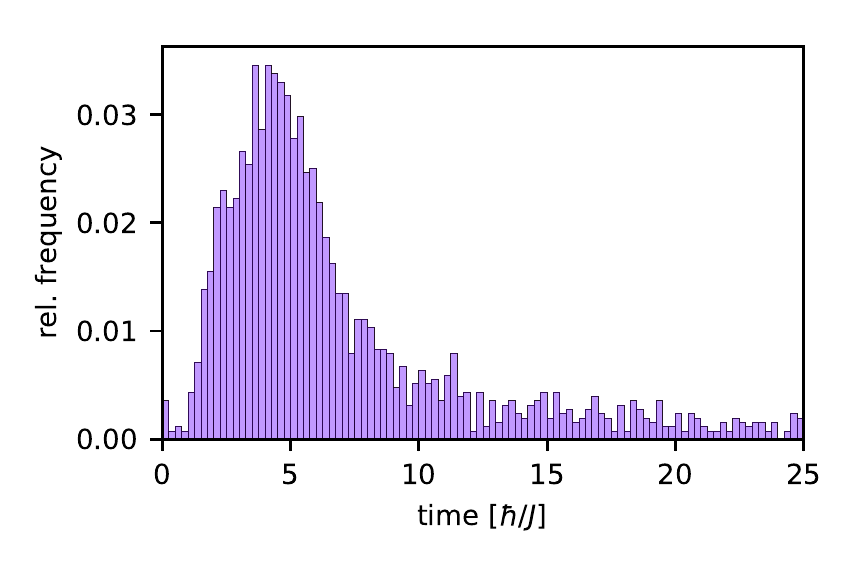}
\caption{Temporal distribution of jumps. The histogram deviates strongly from an exponential behavior, which one would expect for independent jumps in a memoryless system.}
\label{distribution}
\end{figure}

In the experiment, we can observe the system only for a finite time because of atom loss. As a consequence, we typically observe only one jump in a single trajectory (in very rare cases, we observe two). Therefore, we chose two different initial conditions to determine the rates $\gamma_{\mathrm{ns}}$ and $\gamma_{\mathrm{sn}}$ individually, see also Fig.$\,$\ref{setup}. In Fig.$\,$\ref{rates}, we show the resulting effective Liouvillian gap, $\gamma_{\mathrm{ns}}+\gamma_{\mathrm{sn}}$, as a function of the tunneling coupling for fixed dissipation strength. At the minimum, the gap is smaller than one Hertz. This is two to three orders of magnitude smaller than the dissipation rate $\gamma_{\mathrm{diss}}$ and the tunneling coupling $J/\hbar$, which are the intrinsic time scales of the system. The large difference in time scales is a strong indication of a dissipative phase transition and we take the minimum of the relaxation rate to be the position of the phase transition. Around the critical point we find metastability of the two initial states, which manifests itself in the emergence of intermediate hysteresis  for finite measurement times (Fig.$\,$\ref{jumpshare}). Note that only in the thermodynamic limit, which corresponds in our case to an infinite transverse extent of all lattice site, the Liouvillian gap is expected to close completely \cite{Macieszczak2016,Minganti2018}, thus providing bistability at the critical point. The toy model introduced above neglects all density matrices except of $\rho_{\mathrm{ss}}$ and $\rho_{\mathrm{ns}}$ . Therefore, the effective Liouvillian gap is only an upper limit for the true Liouvillian gap. Strictly speaking, the large number of allowed density matrices makes it practically impossible to experimentally measure the true Liouvillian gap. Interestingly, experiment and theory encounter similar constraints in this respect.

So far, our findings are qualitatively similar to those for a single mode nonlinear Kerr resonator. We now turn to the question, what genuine characteristics emerge from the multi-mode nature of our system and make it distinct from the single-mode system.

We start our discussion by analyzing the jumps in more detail. Because we are using interacting bosons, we expect to pass a condensation process during a jump from the normal to the superfluid state. Such a condensation process happens under non-equilibrium conditions, characterized by the competition between the coherent drive, atomic collisions/interactions and dissipation. Information on this non-equilibrium condensation process can be obtained from the time resolved site occupancy during the change and the transverse extent of the atom cloud. Figure\,\ref{avgJump} shows the averaged time-resolved detection signal as well as the transverse extent of the atomic cloud during state changes from the normal to superfluid state. We can identify three different stages. After an initial slow exponential growth of the atom number a sudden rise in the growth rate is apparent. An exponential fit to the initial part of the averaged trajectory is shown to highlight the speed-up in the filling rate above a critical atom number. In a final step, the filling of the central site eventually equals that of the neighboring sites and remains constant. 

It is instructive to look at the timescale of this process by analyzing the duration of the state change in each trajectory individually. We find that the shortest duration for a state change from the normal to the superfluid state is given by the tunneling time, $\tau\approx \hbar/J$ (Fig.$\,$\ref{duration}a). This can be explained as follows. For a small atom number in the central site we only expect an incoherent transport of particles into that site. However, when the density in the central site exceeds a critical value, high mode occupancy builds up. Fluctuations then trigger a condensation process, thus allowing for a coherent Josephson type of transport into the central site. For fully established phase coherence across the central site, the maximum possible coherent transport is expected. For scrambled and fluctuating phases, much longer durations are possible. We observe that the average duration of the state change is about four tunneling times and the longest duration for a state change can be even 20 times larger. In the other direction, from the superfluid to the normal state, the dissipation rate is the determining factor (Fig.$\,$\ref{duration}b). No state change can be faster than the inverse of the dissipation rate and this is also our experimental observation. The average state change is again about two to five times longer and few trajectories show even longer durations. For both types of jumps, such large variations hint at the dominant role of fluctuations for the system dynamics. 

Another important observation is that history appears to be relevant in our system. If the state changes were fully independent of the elapsed evolution time, the temporal distribution of the state changes would follow an exponential decay \cite{probModRoss}. In our experiment, however, the temporal distribution of state changes is non-trivial (Fig.$\,$\ref{distribution}). We find small probabilities of a state change for early times, increasing probabilities as time elapses further followed by a final slow decay of probability. This supports our previous observation in Fig.$\,$\ref{avgJump}, that a state change has preceding internal dynamics. 

To explain the above reported observations for state changes from the normal to the superfluid state, we have developed a simple stochastic rate model, which describes the microscopic system dynamics in the normal state. As indicated in Fig.$\,$\ref{setup}, the mean-field energy (chemical potential) in the neighboring sites renders the direct tunneling process into the ground state of the central site off-resonant. In a single-mode Josephson array, this would lead to macroscopic quantum self-trapping, thus inhibiting any transport \cite{Albiez2005}. The multi-mode nature of our system changes this picture. Atoms are allowed to tunnel into higher orbital states of the central site, where they are subject to collisions between particles in different orbital states. Due to a Franck-Condon overlap between the condensate wave function in the neighboring site and the single particle states in the central site, the effective tunneling coupling is reduced and depends on the atom number in the central site \cite{NDC}. Overall, we get an effective tunneling rate $\Gamma_{\mathrm{eff}}$, which depends on the dissipation rate $\gamma_{\mathrm{diss}}$ and the effective tunneling coupling $J_{\mathrm{eff}}(N)$, where $N$ is the number of atoms in the central site. We thereby assume that the effective tunneling coupling depends linearly on the atom number and reaches the value of $J$ for equal filling with the neighboring sites \cite{NDC}. We find (see methods for further details):

\begin{equation}
\begin{aligned}
    \Gamma_{\mathrm{eff}}(N) &= \frac{J_{\mathrm{eff}}^2(N)}{(\gamma_{\mathrm{diss}} + \gamma_{\mathrm{coll}}) + J_{\mathrm{eff}}^2(N)/\gamma_{\mathrm{diss}}} ,
\end{aligned}
\label{gammaeff}
\end{equation}

where $\gamma_{\mathrm{coll}}$ is the collision rate in the lossy site in the normal state. The stochastic differential equation describing the population $N$ in the lossy site then reads

\begin{equation}
\begin{aligned}
    dN &=  \big(\Gamma_\mathrm{eff}(N) N_\mathrm{F} - \gamma_\mathrm{diss}N\big)dt +  \sigma dW_t ,
\end{aligned}
\label{ratemodel}
\end{equation}
where $N_\mathrm{F}$ is the atom number in the neighboring sites. The term $\sigma dW_t$ with $\sigma=\sqrt{\gamma_\mathrm{diss}}$ describes the fluctuations induced by the dissipation. The dynamics resemble those of geometric Brownian motions (for more details on the model see methods section). Note that due to the $N$-dependence of the effective tunneling rate, the rate equation is nonlinear.

\begin{figure}[t!]
\centering
\includegraphics[]{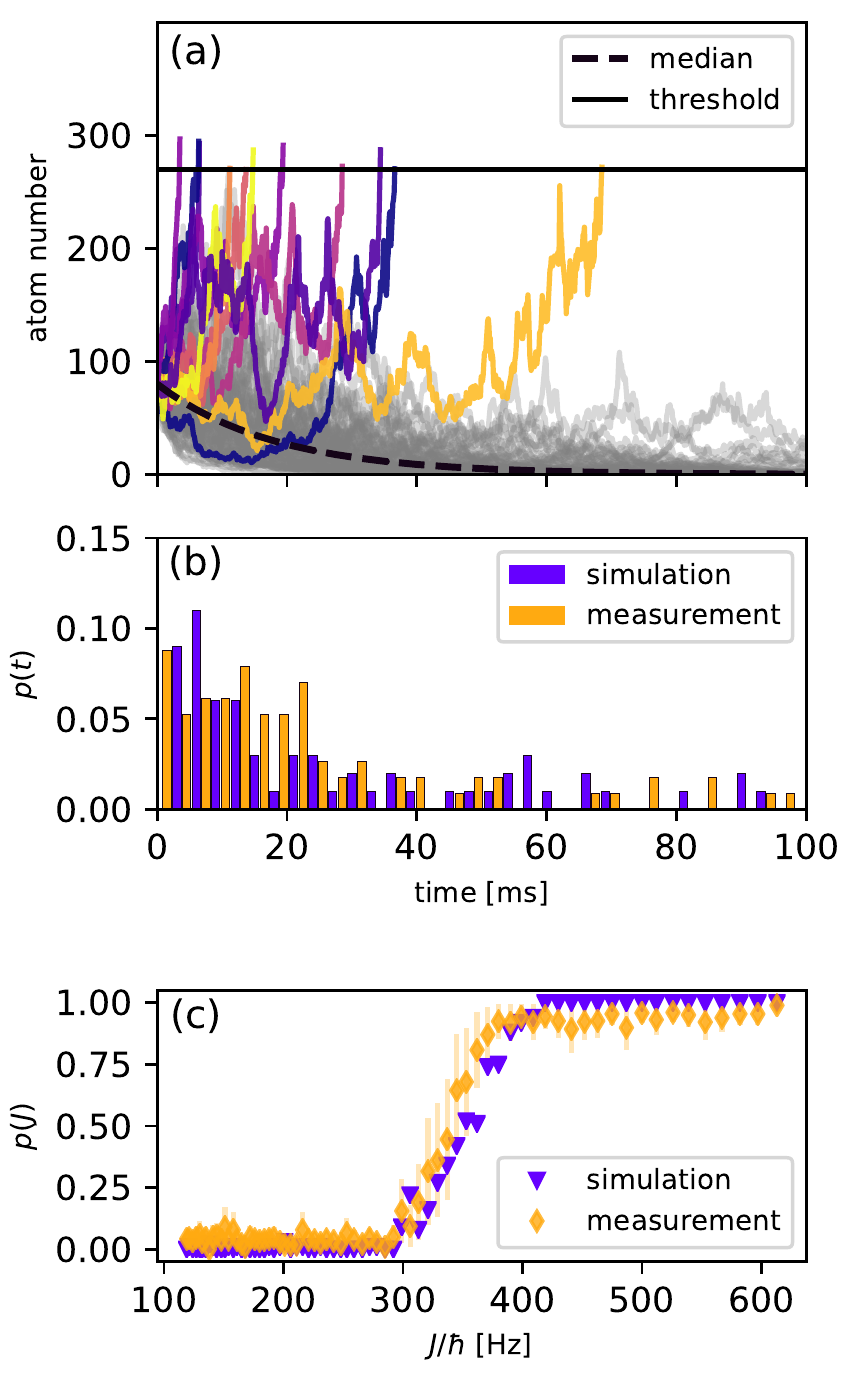}
\caption{Simulation results. (a) $\SI{100}{}$ simulated trajectories initialized in the normal state. While the average trajectory (dashed line) shows an exponential decay, there are nevertheless trajectories ($\SI{8}{}$ colored), that reach a threshold well above the initial number of atoms. (b) The distribution of individual times of the jumps (shown for a tunneling coupling of $J/\hbar = \SI{380}{Hz}$) is reproduced as well as (c) the fraction of trajectories that exhibit a jump.
Error bars show the standard deviations.}
\label{compare_sim}
\end{figure}

To simulate the system we time evolve many individual trajectories by Monte-Carlo simulations. When a trajectory reaches a threshold filling for the first time macroscopic phase coherence builds up and the jump occurs. In Fig.$\,$\ref{compare_sim}a, we show a series of simulated trajectories. We take the time of the jump to be the time the threshold is reached. The introduction of a threshold filling followed by a condensation process gets backing from the measurement of the width of the atomic cloud within the central site during the jump (Fig.$\,$\ref{avgJump}): All atoms tunneling from a neighboring reservoir into the central site have total energy equal to the chemical potential $\mu$ of the neighboring sites. In the beginning of the refilling process, when the central site contains barely any atoms, the interaction energy is converted into kinetic energy, leading to an effective finite temperature. From an approximation of the density distribution of the thermal part inside the central side by a Gaussian function the width of the cloud in the central site is calculated to be $\sigma_{\mathrm{th}} = \sqrt{k_B T/\left(m \omega^2\right)}$. Here $\omega$ denotes the radial trap frequency, $k_B$ the Boltzmann constant, $m$ the mass of rubidium atoms, and $T$ the temperature in the site. The condensate's radial extent in the central site is approximated by the Thomas-Fermi radius $r_{\mathrm{TF}} = \sqrt{\mu/\left(2 m \omega^2\right)}$. The calculated widths for our system are $\SI{2.1\pm0.1}{\upmu\meter}$ for the normal/thermal gas and $\SI{1.7\pm0.1}{\upmu\meter}$ for the condensate. Thus, the transverse extent of the cloud shrinks with increasing atom number. This is a consequence of the condensation process and the reduction of thermal fluctuations. The experimental results (indigo line in Fig.$\,$\ref{avgJump}) are in good agreement with this prediction. 

In Fig.$\,$\ref{compare_sim}b we compare the temporal distribution of the simulated jumps with our measurement and in Fig.$\,$\ref{compare_sim}c we show the comparison for the fraction of trajectories exhibiting a jump. The only fitting parameter of our model is the threshold filling and we find a critical value of $\SI{300}{}$ atoms, which is in good agreement with the value extracted from the measurement (Fig.$\,$\ref{avgJump}).
 
This has to be compared to the critical atom number for Bose-Einstein condensation in thermal equilibrium. 2D BEC theory \cite{Fletcher2015} gives a value of $\SI{100}{}$ atoms. This confirms that the fluctuations increase the critical atom number for condensation in the central site and underlines the qualitative difference between a dissipative phase transition and a phase transition in thermal equilibrium.

In summary, we have experimentally characterized a first-order dissipative phase transition in a multi-mode many-body quantum system. Our results extent the experimental studies on nonlinear optical Kerr resonators. While some features are qualitatively similar, the microscopic dynamics are much richer due to the multi-mode nature. For the future, we expect to see more experiments working on related systems, such as the driven dissipative 2D Bose-Hubbard model or corresponding fermionic counterparts. For the system at hand, a detailed study of the scaling of the Liouvillian gap with the system size would allow for a finite size scaling towards the thermodynamic limit. The measurement of critical exponents in the vicinity of the phase transition would allow for a deeper understanding of the classification of the phase transition \cite{Casteels2017}.

\section{methods}

{\bf The master equation}\\

The master equation of our system is constructed in the following way. We first note that the atoms occupy the lowest band along the lattice direction. We are well within the tight binding approximation along this direction and introduce the discrete site index $i$. The central site corresponds to $i=0$. The kinetic energy in the lattice direction results in the tunneling coupling $J$. In the two transverse directions ($x,y$) each site experiences a harmonic confinement with radial frequency $\omega_{\perp}$. The Hamiltonian of the central site  in terms of bosonic field operators ($\hat{\Psi}_0(x,y)=\hat{\Psi}_0$) reads 

\begin{align}
\hat{H}&=\int dx dy \hat{\Psi}^\dagger_0\left[-\frac{\hbar^2}{2m}\nabla_{\perp}^2 + \frac{1}{2}m\omega_{\perp}^2 (x^2+y^2) \right] \hat{\Psi}_0 \notag\\
& + \frac{g_{\mathrm{2D}}}{2}\int dx dy\hat{\Psi}^\dagger_0\hat{\Psi}^\dagger_0 \hat{\Psi}_0\hat{\Psi}_0\\ 
& - J \int dxdy \left(\Phi^*(x,y)e^{i\mu t /\hbar} \hat{\Psi}_0 +  \Phi(x,y)e^{-i\mu t /\hbar} \hat{\Psi}^\dagger_0\right) \notag .
\end{align}

The effective interaction strength in the lattice site is given by $g_{\mathrm{2D}}$. The coherent drive is given by a mean-field wave function of the condensate in the two neighboring sites given by $\Phi(x,y)e^{-i\mu t /\hbar}$, where $\mu$ is the chemical potential. The large number of additional adjacent reservoir sites justifies to set $\Phi(x,y)$ constant in time.

The losses are included via a master equation, 

\begin{align}
\dot{\hat{\rho}}&=\hat{\mathcal{L}}\hat{\rho} \notag\\
&=-\frac{i}{\hbar}\left[\hat{H},\hat{\rho}\right]\\
&+ \frac{\gamma_\mathrm{diss}}{2} \int dx dy \left(2\hat{\Psi}_0 \hat{\rho} \hat{\Psi}^\dagger_0 -  \hat{\Psi}_0^\dagger\hat{\Psi}_0 \hat{\rho} - \hat{\rho}\hat{\Psi}_0^\dagger\hat{\Psi}_0 \right) \notag .
\end{align}

As outlined in the main text, an exact theoretical treatment is not possible. A stochastic mean-field description of this master equation has been developed and analyzed in detail in Ref.\,\cite{Reeves2021}.\\

{\bf Experimental setup}\\

We use a Bose-Einstein Condensate (BEC) of ${}^{87}$Rb containing about $\SI{150e3}{}$ atoms. The atoms are trapped in a focused CO${}_{2}$-laser beam with a beam waist of $\SI{34}{\upmu\meter}$. The condensate is cigar-shaped with dimensions $(115 \times 6\times 6)\SI{}{\upmu m}$ in an approximately harmonic trap with frequencies $2\pi\cdot(9.8, 197, 183)\SI{}{Hz}$. After its preparation the BEC is loaded into a one-dimensional blue detuned optical lattice created by two beams ($\lambda=\SI{774}{nm}$, beam waist $\SI{500}{\upmu m}$) intersecting at $90^{\circ}$. The resulting lattice has a period of $d = \SI{547}{nm}$. Each site contains a small 2D BEC with about $\SI{800}{}$ atoms at the center of the trap. The total number of lattice sites is about $\SI{200}{}$. In addition, an electron column implemented in the vacuum chamber provides a focused electron beam which is used to introduce a well-defined local particle loss as a dissipative process at one site of the lattice. In order to ensure a homogeneous loss process over the whole extent of this lattice site we rapidly scan the electron beam in the transverse direction ($\SI{3}{kHz}$ scan frequency) with a sawtooth pattern. To adjust the dissipation strength $\gamma$, we vary the transverse extent of the scan pattern. In the experiments considered in this paper we start out with a BEC, load it into the 1D lattice at a depth of $s=30$ (we give the values of the lattice depth $V_0 = s\,E_\mathrm{rec}$ in units of the recoil energy $E_\mathrm{rec} = \pi^{2}\hbar^{2}/\left(2 m d^{2}\right)$) and - depending on the initial conditions - deplete one site. We then reduce the lattice depth to a value between $s=6$ and $s=12$ while keeping the electron beam scanning over the depleted site. Our measurement signal is the time-resolved ion signal from the scan at the lower lattice depth. Only in the case of the dynamical hysteresis measurement, the lattice depth is varied in time (for the measurement presented in Fig.$\,$\ref{jumpshare}a the lattice depth was varied at a constant rate from $s=4$ to $s=20$ and back).
\newline

{\bf Microscopic model}\\

In order to develop a rate model for the dynamics in the central site in the normal state (low filling), we start out from a two-level system. The neighboring site corresponds to the ground state. Coherent coupling to a radially excited state in the central site is provided by an effective tunneling coupling $J_\mathrm{eff}$, which takes into account the finite Franck-Condon overlap between the neighboring reservoir site and the radially excited state \cite{Bistability}. Atom loss from this radially excited state is decribed by the dissipation rate $\gamma_{\mathrm{diss}}$ and dephasing caused by atomic collissions is included via $\gamma_{\mathrm{coll}}$. The rate equation for the central site is then given by 

\begin{equation}
\label{eq:2rate}
    \frac{dN}{dt} = \Gamma_{\mathrm{eff}} N_{\mathrm{F}} - \gamma_{\mathrm{diss}}N,
\end{equation}

where $N$ is the number of atoms in the central site and $N_F$ is that of the neighboring sites. Due to the large number of reservoir sites to which the neighboring site is coupled we set $N_F$ as constant. The effective tunneling rate $\Gamma_\mathrm{eff}$ in this effective two-level system is then given by

\begin{equation}
\label{eq:effrate}
    \Gamma_{\mathrm{eff}} = \frac{J_{\mathrm{eff}}^2}{(\gamma_{\mathrm{diss}} + \gamma_{\mathrm{coll}}) + \frac{J_{\mathrm{eff}}^2}{\gamma_{\mathrm{diss}}}}.
\end{equation}

In our system atoms can tunnel between the full site ($\ket{\Phi_{\mathrm{F}}}$) and the resonant mode ($\ket{\Phi_{\mathrm{E}}}$) of the empty site with the effective tunnel rate $\Gamma_{\mathrm{eff}}$. Introducing the overlap integral $\eta = \vert\braket{\Phi_{\mathrm{F}}\vert\Phi_{\mathrm{E}}}\vert$, and assuming a linear dependence on the atom number for finite filling, we rewrite the effective tunnel as follows \cite{NDC}:

\begin{equation}
\label{eq:effratefinal}
    \begin{aligned}
    \Gamma_{\mathrm{eff}} &= \frac{\Big(\frac{N}{N_{\mathrm{F}}} (J-\eta J) + \eta J\Big)^2}{\Big(\gamma_{\mathrm{diss}} + \gamma_{\mathrm{coll}}(N)\Big) + \frac{\Big(\frac{N}{N_{\mathrm{F}}} (J-\eta J) + \eta J\Big)^2}{\gamma_{\mathrm{diss}}}}.
    \end{aligned}
\end{equation}

For low filling in the central site, as it is the case in the normal state, we linearize the above expression:

\begin{equation}
\label{eq:rateNfinal}
    \begin{aligned}
    \dfrac{dN}{dt} &= \lambda N + a
    \end{aligned}
\end{equation}

To model the fluctuations induced by the dissipation we introduce a stochastic term and obtain the stochastic differential equation
\begin{equation}
\label{eq8}
    dN =  \left(\lambda N + a\right) dt + \sigma dW_t
\end{equation}
with a Wiener process $W_t$ and
\begin{equation}
\label{eq:sigma}
    \begin{aligned}
        \sigma &= \sqrt{\gamma_{\mathrm{diss}}}.
    \end{aligned}
\end{equation}

To solve Eq.$\,$\eqref{eq8} we use the Itô formula, a variant of the chain rule for stochastic multi-variable functions \cite{Ito1944,Ito1946,StchInt}. The solution to this SDE is then given by
\begin{equation}
    \label{SDE_solution}
    N(t) = \Big(N(0)+c\Big) e^{(\lambda-\frac{1}{2}\sigma^2)t + \sigma \sqrt{t}x} - c
\end{equation}
with a standard normal variable $x$ and $c = a/\lambda$.

\section{Data availability} 
The data that support the findings of this study are available from the corresponding author upon reasonable request. 

\section{Acknowledgements}

We gratefully acknowledge fruitful discussions with M. Davis, M. Fleischhauer, C. D. Mink, A. Pelster, and M. Reeves. This project was funded by the DFG (German Science Foundation) within the CRC TR185 OSCAR project number 277625399. We thank A. Gil Moreno for assistance in the early stage of the experiment.

\section{Author Contribution}

J.B., C.B. and M.R. performed the measurements. J.B. and M.R. evaluated and analyzed the data. H.O. has supervised the experiment. J.B. wrote the manuscript. All authors discussed the results and contributed to the manuscript. 

\section{Competing interests}
The authors declare no competing interests.

\bibliography{dissipativePhaseTransition}

\end{document}